\documentstyle[aaspp4,12pt]{article}  
\begin{document}
\title{\bf 
Wavelength Doesn't Matter: Optical vs. X-ray Luminosities of Galaxy Clusters}

\author{Christopher J. Miller\altaffilmark{1},
Adrian L. Melott\altaffilmark{2,}\altaffilmark{3}, and Robert C. Nichol\altaffilmark{2}}
\altaffiltext{1}{Department of Physics \& Astronomy, University of Maine, Orono, ME 04469}
\altaffiltext{2}{Department of Physics, Carnegie Mellon University, Pittsburgh, PA 15213}
\altaffiltext{3}{Department of Physics \& Astronomy,University of Kansas, Lawrence, KS 66045}

\begin{abstract}

We examine here the relationship between the total X-ray and
optical luminosities of groups and clusters of galaxies taken from
various samples in the literature.  The clusters and groups were drawn
from four different catalogs: (1) the Abell/ACO catalog, (2) the
Edinburgh-Durham Cluster Catalog, (3) the RASS Bright Cluster Sample,
and (4) galaxy groups selected from the CfA redshift survey. These
catalogs represent a significant cross-section of cluster selection
techniques as well as a wide range in mass scale and can be considered
statistically independent. We have calculated new cluster X-ray
and optical luminosities if not already
available in the literature. Based on 126 systems, our
analysis shows that the total optical luminosity of a cluster is
directly proportional to the total X-ray luminosity over a wide mass
range and across all four cluster samples studied herein.  We also
show that  total cluster optical luminosity is a good indicator of
cluster virial mass, whereas richness is not.
Our results suggest that (1) the selection
method of galaxy clusters may not be crucial, providing wider latitude in
assembling catalogs, (2) the luminosity per baryon does not vary
systematically between systems ranging from groups to very rich
clusters.  We propose that future optical catalogs of
clusters use the total optical luminosity of a cluster, instead of
galaxy richness,  as luminosity appears to be a better measure of
cluster mass and can be directly related to X--ray catalogs of
clusters. This will also facilitate the first direct comparison
between the growing number of X-ray and optical catalogs.
\end{abstract}

\section{Introduction}

Clusters of galaxies are the most massive gravitationally bound and
collapsed objects in the Universe and are important systems
for studies of the formation and evolution of
Large-Scale Structure (LSS) in the Universe. Their large
mean spatial separations ($\sim 60-100h_{50}^{-1}$Mpc) make them
excellent tracers for LSS. The cosmological
subfield of LSS is still in its infancy, with new and exciting
discoveries occurring often.  As large, all-sky surveys continue to
catalog the heavens, galaxy clusters will remain the most efficient
tool for mapping LSS. The importance of a complete map of LSS has
become evident in some recent discoveries of how the internal
properties of clusters can be affected by surrounding large-scale structure
(Novikov et al. 1999; Loken et al. 1999; Miller et al. 1999).
Additionally, clusters are often thought to be a ``fair sample'' of
the mass in the Universe, and therefore useful for estimating the
relative fractions of baryonic and ``dark'' matter (White {\it et al.}
1993; White and Fabian 1995; however see Qin and Wu 2000).

Most of the largest cluster surveys presently available in the
astronomical literature have been constructed at optical wavelengths
using galaxy population overdensities to define the individual
clusters. The best known example of this technique is the Abell (1958)
and Abell, Corwin, and Olowin (1989- hereafter ACO) optical cluster
catalogs. 
Many researchers have discovered observational
selection biases within these catalogs as a direct result of
the criteria used in their creation (see {\it e.g.}  Sutherland 1988;
Efstathiou 1992). 
One of the more deserved critiques of the Abell/ACO
catalogs is the inaccuracy of the the richness classifications 
(Lumsden et al. 1992; van Haarlem {\it et al.} 1997; White, Jones and Forman 1999).  Richness has long
been used as a classification in nearly all optically selected cluster
catalogs -- {\it i.e.} Abell, APM, EDCC, PDCS etc -- yet it is rather
unphysical to quantify cluster properties (such as mass, size, etc)
based on a simple two--dimensional galaxy count within an arbitrary
radius, to an arbitrary magnitude-limit with a local and/or global
field galaxy count subtracted off.  This ignores a
significant amount of information contained in the individual galaxies
(such as intrinsic brightness and color).  Also, the richness of a cluster
can be severely contaminated by projection effects as one goes
to higher redshift.

In addition to these problems with galaxy richness, the
recent discovery of ``fossil groups'' and ``dark clusters'' further
illustrate the deficiencies of richness.  Both Vikhlinin et al. (1999)
and Romer et al. (1999) have discovered several examples of ``fossil
groups'' which are the proposed relics of group formation and
evolution {\it i.e.} all the galaxies in the group have `cooled' and
fully merged into one large galaxy in the center of the potential
well. However, the X--ray emitting gas is still `hot' and appears
extended. Therefore, in an optical survey these systems appear to be
one galaxy while in the X--rays they appear like groups of
galaxies. Clearly there is a dark--matter halo associated with such
systems which would be completely missed in a classical,
richness--limited survey. However, in an optical luminosity--based cluster
survey, these systems would be included since the one central galaxy is
typical many $L^{\star}$ in brightness (see Romer et al. 1999).

\section{The Cluster Samples}

We propose here dropping the notion of cluster richness
in favor of  total optical luminosity of a cluster; This can either
be in a single passband or as a function of color {\it e.g.}
constraining the color of the light to that appropriate for elliptical
galaxies.  Thus we allow the individual galaxy properties 
to re-enter the analysis. Currently, a large portion of
the sky has reasonably accurate galaxy magnitudes through the
APM/COSMOS galaxy catalog. This database covers the Southern galactic
cap. There is a significant amount of work being done in the Northern
hemisphere ({\it e.g.}  the DPOSS survey (Djorgovski {\it et al.}
1999) and the Sloan Digital Sky Survey). Soon, we will have precise
magnitude, and color information for most bright ($b_j \le 20.5$) galaxies
throughout the entire sky. We show in this work that future cluster
identifications would be better quantified by total
optical cluster luminosity than by richness.

\subsection{The Edinburgh-Durham Cluster Catalog}

The Edinburgh-Durham Cluster Catalog (Lumsden et al. 1992 -hereafter
EDCC) was objectively selected from the Edinburgh/Durham Southern
Galaxy Catalogue (EDSGC; Collins, Nichol \& Lumsden 1992, 2000) which
was built from machine-scans of 60 UK Schmidt photographic survey
plates.  In total, the EDCC contains over 700 galaxy overdensities
covering over $1000{\rm deg^2}$ of the sky centered on the South
Galactic Pole. The reader is referred to Lumsden {\it et al.} (1992)
for details about the EDCC.
We supplemented the optically--selected EDCC catalog with X--ray
information taken from the ROSAT All-Sky Survey (RASS) Bright Source
Catalog (Voges {\it et al.} 1999).  This was achieved through cross--correlation of
the EDCC centroids with the RASS--BSC positions. In Figure 1, we show
our separation analysis for our EDCC--BSC sample which demonstrates a
positive correlation (above that expected from random) for separations
of less than 4 arcmins.  Below 4 arcminutes, we would only expect a
random 5.3 matching pairs while we see 53 matching pairs. Therefore,
we use 4 arcminutes as our matching radius which should ensure that
90\% of our these 53 EDCC clusters have a true X--ray companion.  We
then cut the EDCC--BSC sample to include clusters with a known
redshift (Collins et al. 1995) as well as those EDCC clusters
coincident with an extended BSC source, thus guaranteeing minimal
contamination from active galactic nuclei (AGN).  These restrictions
leave us with a subset of 20 EDCC--BSC clusters from the original 53
matches.

The X-ray luminosities of the EDCC--BSC clusters were computed as
follows. First, the observed BSC count rate (counts per second) was
converted to a flux (ergs per second per ${\rm cm^2}$) by integrating
a thermal bremsstrahlung spectrum of ${\rm T_e}=5$keV and metallicity
of 0.3 solar over the ROSAT PSPC response function
($0.1\rightarrow2.4$keV) and comparing this to the observed count rate
(we corrected for absorption by the galactic neutral hydrogen using 
the data of Stark et al. 1990). Second, we
corrected this aperture flux into a total cluster flux using a
standard King profile ($r_c=250h^{-1}$kpc and $\beta=0.66$).
Finally, our fluxes were converted into X-ray
luminosities using
$q_o=0.5$, and $h_{50} = H_0/50$ (although we note that the choice of
$q_o$ makes little difference in our calculations).

The optical luminosities for these 20 EDCC clusters were 
calculated using photometry from the EDSGC
which was calibrated 
to an accuracy of $\Delta m\simeq0.1$ across the whole survey (see
Collins et al. 2000).
We first calculated the absolute magnitudes for all galaxies with
$b_j \le 20.5$ within each cluster using the same
methods as Lumsden {\it et al.} (1992). We use
$K(z) = 4.14z - 0.44z^2$ as the $K$-correction suitable for the
$b_j$ passband (Ellis 1983), and $A(b_j)$ as the extinction values taken from
the Schlegel, Finkbeiner, and Davis (1998) reddening maps. 
Assuming that each galaxy lies at the distance of the cluster, we
summed the individual galaxy luminosities to determine the local
average background luminosity out to $20h_{50}^{-1}$Mpc around the 
cluster center. We then subtracted the local background
from total cluster luminosity as determined within $2h_{50}^{-1}$Mpc
of the cluster center. 
We applied apparent and absolute magnitude constraints
and found only small ($\sim 10\%$) variations in our results. We also
calculated the background using the average luminosity within the
ring from $10h_{50}^{-1}$Mpc to $20h_{50}^{-1}$Mpc, effectively excluding the
optical emissions from the clusters. Again, our final results only
varied by $\sim 10\%$. From these analyses, we conclude that our methods
are robust and we measure errors on $L_o$ according to the variation
in the choice of absolute magnitude limits (from $-24 < M_{b_j} < -21$).  
We present our optical and X-ray luminosities in Table 1.

In this work, we are studying a unique subset of the EDCC
data, {\it i.e.} only those that have a redshift and extended X-ray
emission in the RASS--BSC. Therefore, the sample presented in Table 1
has a complicated selection function since we have made two
flux cuts (for the EDCC \& BCS), a cut on X--ray extent and a cut on
richness (since the high richness clusters have measured redshifts;
Nichol et al. 1992). However, this selection function should not
effect our analysis and results since we are simply using this sample
to study the relationship of optical--to--X--ray luminosities of these
systems and are presently uninterested in the cross--comparison of
these qualities between clusters.

\subsection{The Abell/ACO Cluster sample}

The second sample we have used is the Abell/ACO subset presented by Fritcsh
\& Buchert (1999, hereafter FB) who used a sample of 78 Abell/ACO
clusters to create a fundamental plane in $L_{optical}$, $L_{X-ray}$
and half-light radius and $R_{optical}$. They determined their optical
luminosities after subtracting a local galaxy/photon distribution and
applying a Schechter luminosity function with $M_* = -21.8 $ and
$\alpha = -1.25$. The individual galaxy magnitudes were taken from the
COSMOS galaxy catalog. The X-ray luminosities were determined from
ROSAT data using a Raymond-Smith code. More details can be found in
FB. Unfortunately, FB do not provide the uncertainties on
their data.

\subsection{The RASS Bright Sample}
Our third cluster sample is the ROSAT All-Sky Survey (RASS) Bright
Sample which contains 130 clusters constructed from the ESO Key
Program.  This has since become the REFLEX cluster survey 
(Guzzo et al. 1999; Bohringer et al. 1998).  The RASS Bright
Sample is an X-ray selected cluster catalog.  Extended X-ray sources
in the RASS data were searched for over-densities in the galaxy
distribution.  This survey is count-rate limited in the ROSAT hard
X-ray band. The RASS Bright Sample covers 2.5 steradian around the
Southern Galactic Cap.  We find 20 RASS clusters within a similar
section of the sky as the EDSGC. However, due to the significantly
different cluster selection and X-ray identification techniques 
between the RASS and
EDCC surveys, we find only five clusters
that are common to both. In other words, although the EDCC and RASS samples
are in the same portion of the sky, the two samples remain nearly
statistically independent. 
The X-ray luminosities for the RASS clusters are listed in 
De Grandi et al. (1999).  We measured optical luminosities for the
RASS clusters similarly to the EDCC clusters (see above) and present them 
in Table 1.

\subsection{Galaxy Groups}
Finally, we use data provided by Mahdavi {\it et al.} 1997 for poor
galaxy groups. Mahdavi {\it et al.} studied 36 groups with at least
five galaxy members selected from the Center for Astrophysics Redshift
Survey (Ramella {\it et al.} 1995).  Nine of these groups were found to
have definite X-ray emissions via RASS data.  The optical properties
were determined from Zwicky magnitudes listed in Ramella {\it et
al.} and Mahdavi {\it et al.}  fit Schechter luminosity functions (with
$\alpha =1$ and $M^*_B = -20.6$) to determine total optical
luminosities to a limit of $M_B=-18.4$.  We note that three groups in
this sample are also $R =0$ Abell clusters (although none are in the
FB data as well).

\section{Results}

In Figure 2, we plot $log_{10} L_x$ versus $log_{10} L_o$ for the Abell/ACO clusters (Fig 2a),
EDCC clusters (Fig 2b), galaxy groups (Fig 2c) and RASS clusters (Fig 2d). 
A correlation analysis
indicates that the optical and X-ray luminosities are linearly related to 
4.4$\sigma$, 3.1$\sigma$, 2.1$\sigma$, and 1.9$\sigma$ significance levels respectively.
[Note: if we remove the three ``outliers'' in the RASS sample, the significance rises
to 3.2$\sigma$. This suggests that these points may not be clusters, but extended AGNs]. 
In all but the group sample, we also provide a robust
best-fit line through the data points.  The fit for the galaxy group
is given by Mahdavi {\it et al.} (1997). 
In all cases, the slopes (listed in Fig. 2) are very close to unity,
suggesting a simple proportionality survives the many procedural
differences in the way these samples were selected and analyzed:
{\it i.e.} Schechter functions were fit to some clusters, while for
others, a simple sum of the individual galaxy luminosities was
performed.  Also, different optical wavelength windows were used {\it
i.e.} $b_j$ for Abell/ACO, EDCC, RASS, and $m_{Zwicky}$ for the
galaxy groups.  Different aperture sizes within which magnitudes were
summed were used: variable radius for Abell/ACO clusters,
$2.0h_{50}^{-1}$Mpc for EDCC clusters and RASS clusters, and
$0.4h_{50}^{-1}$Mpc for the galaxy groups.

At this point, it would be helpful to understand the role of the
amplitude in Figure 2. However, we simply do not have enough
information to constrain any more parameters than the slope.
Some of the effects on the amplitude are more easily quantified,
such as the solar luminosity used in the $L_o/L_{\odot}$ normalization
or the Hubble constant used. The other procedural differences within
these samples are more difficult to ascertain and correct for,
such as the counting-radius used, magnitude-limits, etc.
Therefore, at this point we would like to stress the importance
of the simple proportionality between optical and X-ray cluster
luminosities. 

\section{Discussion}

The simplest possible assumption about the assembled contents and
their emitted radiation seems to produce the observed results.  This
need not have been true.  Optical luminosity measures the total
optical radiation of all the stars in the clusters--which depends on
the efficiency of star formation, the initial mass function of the
stars formed, and the age of the populations.  The X-ray emission
depends on the mass, temperature, and the (square of the) density of
the hot gas.  We have studied a wide range of objects from galaxy
groups up through rich clusters--several orders of magnitude in mass
and luminosity.  The trend which forms the basis of our result
continues through this mass hierarchy.

This result could be the endpoint of a cosmic conspiracy of canceling
effects.  However, a simpler explanation is that ``averaging'' is
effective, even on small scales.  If systems as small as galaxy groups
sample the stellar and gaseous phases of baryonic matter in the
Universe as well as rich clusters, and the overall efficiency and mass
range of star formation is about the same, and if the radiative
efficiency of bremsstrahlung radiation is again about the same on
average, then our result will follow.  A slightly puzzling aspect of
this result is that the larger systems are known to have higher
velocity dispersions--and one would expect higher gas
temperatures. One would think this would lead to greater X-ray
luminosity per unit mass of gas.

Besides the cosmological implications, we must also consider how these
results can be immediately applied in the construction of future
cluster catalogs. With a proper normalization to the $L_x \propto L_o$
relation, we can very quickly and easily estimate cluster X-ray
luminosities and masses using only the galaxy magnitudes and a minimal
amount of spectroscopic and X-ray information.  We need not perform
extensive spectroscopic or X-ray observations of an entire cluster
sample in order to analyze mass functions or correlation-functions and
their dependence on cluster masses and X-ray luminosities.

As an example, we plot $M_o$ vs $L_o$ for the Abell/ACO clusters in
Figure 3.  Here, we have used cluster virial masses ($M_{CV}$) as
published in Girardi {\it et al.} (1998) and the richness counts from
the ACO catalog. The optical luminosities presented in FB were
calculated within radii that included all of the light above the
background. Therefore, we use those masses corresponding to the
largest radii as published in Table 3 of Girardi {\it et al.}  Figure
3 shows that optical luminosity linearly traces
the total mass in galaxy clusters over a wide range, whereas we see a
much weaker dependence of richness on mass. 
Therefore, $L_o$
is much easier to correlate to mass than richness.
Smail {\it et al.}
(1997) have used more distant clusters to show that $M/L$ varies little 
over a much smaller range of mass.
We cannot provide any quantitative limits on $M/L$ for clusters without
more information on the FB optical luminosities.

In this paper, we have demonstrated that the total optical luminosity
of clusters and groups of galaxies is correlated with both the X--ray
luminosity and mass of these systems. This result is very important
because, in the near future, there will be many new surveys of
galaxies and clusters, {\it e.g.} DPOSS, REFLEX, SDSS, EIS, 2MASS {\it
etc.}, and we will need objective methods to a) compare these
different catalogs and b) relate the cluster properties to physically
meaningful quantities.  Optical cluster luminosities will be easy to
compute from these digital catalogs and will allow us to directly
relate optical cluster catalogs with X--ray--selected catalogs ({\it
e.g.} REFLEX, EMSS, SHARC). Moreover, it is hoped that the optical
luminosity of a cluster is more easily related to theoretical cluster
research than the galaxy richness, since we have shown that the optical
luminosity is simply probing the density in baryons in the cluster. It
is appears that galaxy richness has a similar physical motivation.

\begin{figure}
\plotone{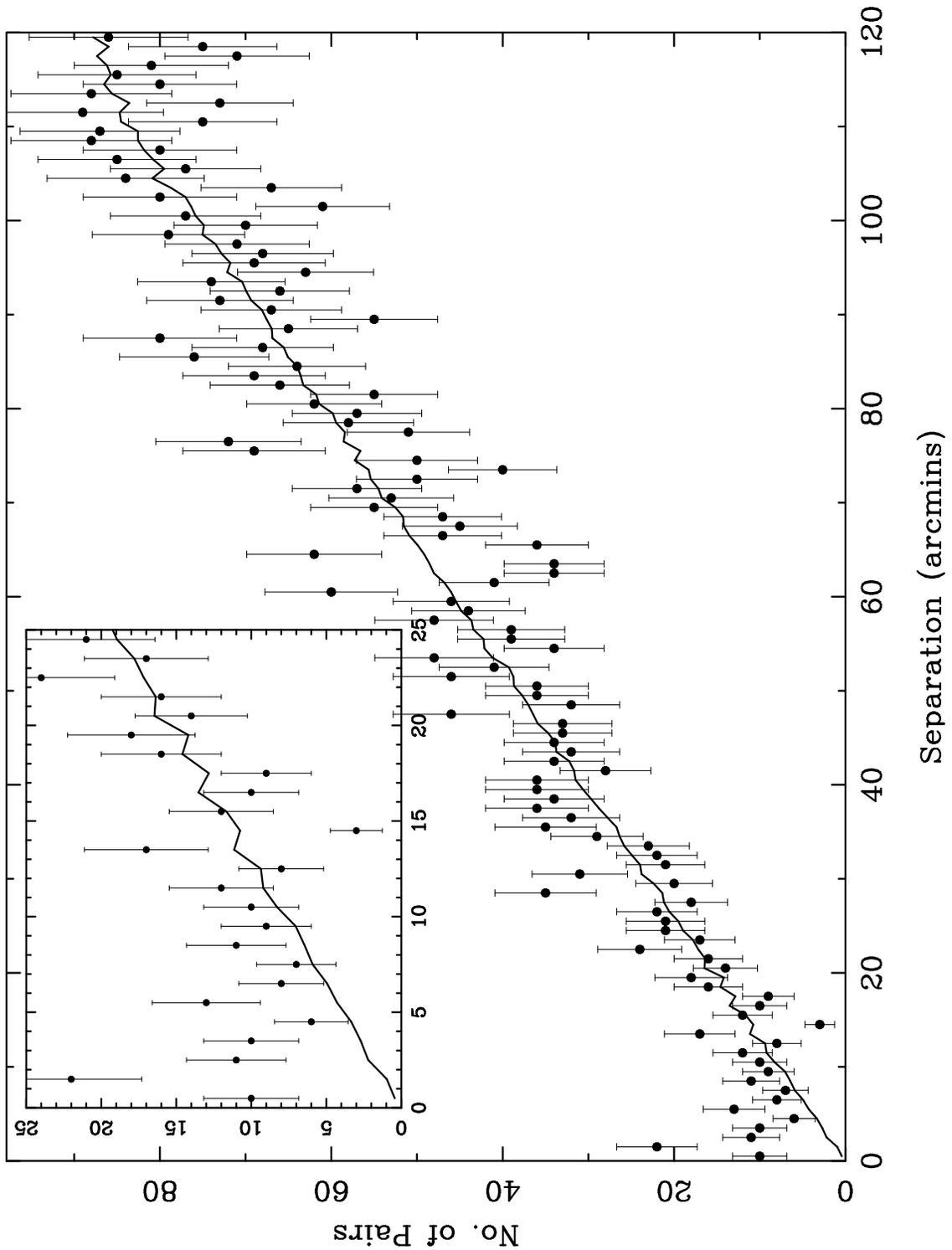}
\caption[]{
EDCC--BSC pair counts as a function of angle. 
A random sample would produce only 5.3 matching pairs below 4 arcmins
while we see 53 matching pairs. Therefore, we use 4 arcminutes
as our matching radius which should ensure that 90\% of our
EDCC clusters have a true X--ray companion.}
\end{figure}

\begin{figure}
\plotone{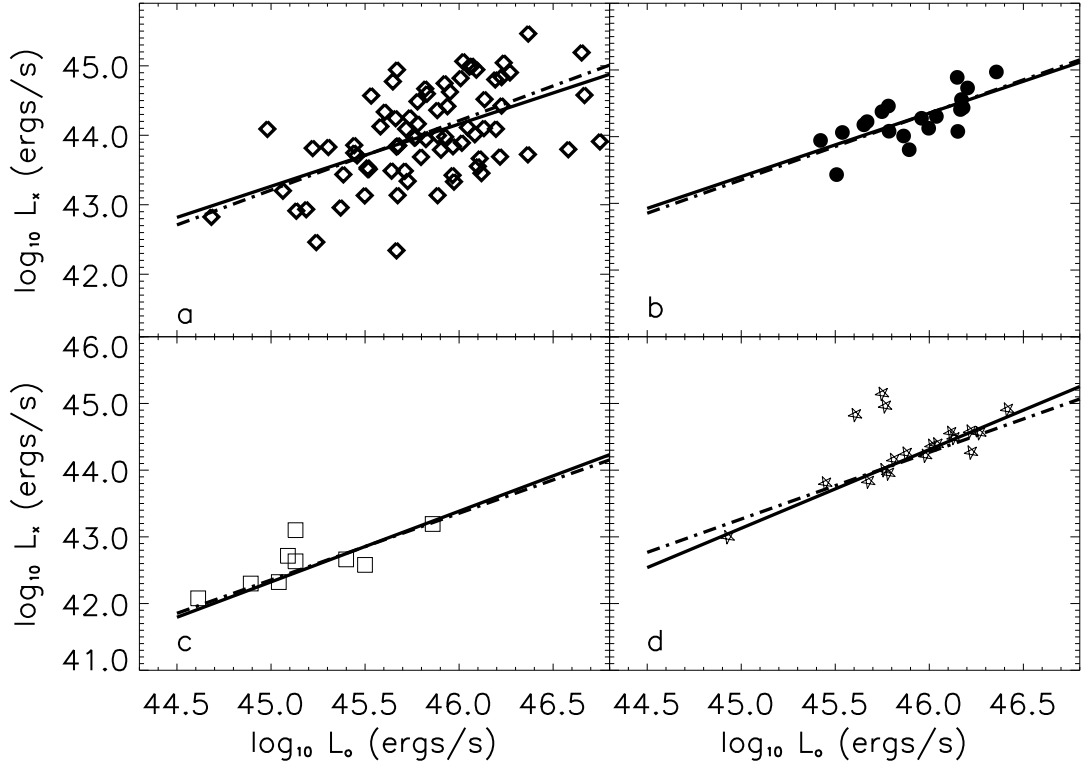}
\caption[]{We plot log$L_o$ vs. log$L_x$ for the samples defined
in Section 2. The error bars are excluded for reasons of
clarity (although they are listed in Table 1).
In each plot, the solid line corresponds to best-fit
to the data. {\bf (a)} is the Abell/ACO sample where $logL_x \propto logL_o^{0.90\pm{0.17}}$.
{\bf (b)} is the EDCC sample with  $logL_x \propto logL_o^{0.95\pm{0.22}}$. {\bf (c)} is
the CfA group sample with  $logL_x \propto logL_o^{1.06\pm{0.11}}$. {\bf (d)} is the
RASS sample with  $logL_x \propto logL_o^{1.18\pm{0.08}}$. In each plot, the dashed-line
corresponds to $logL_x \propto logL_o$.}
\end{figure}

\begin{figure}
\plotone{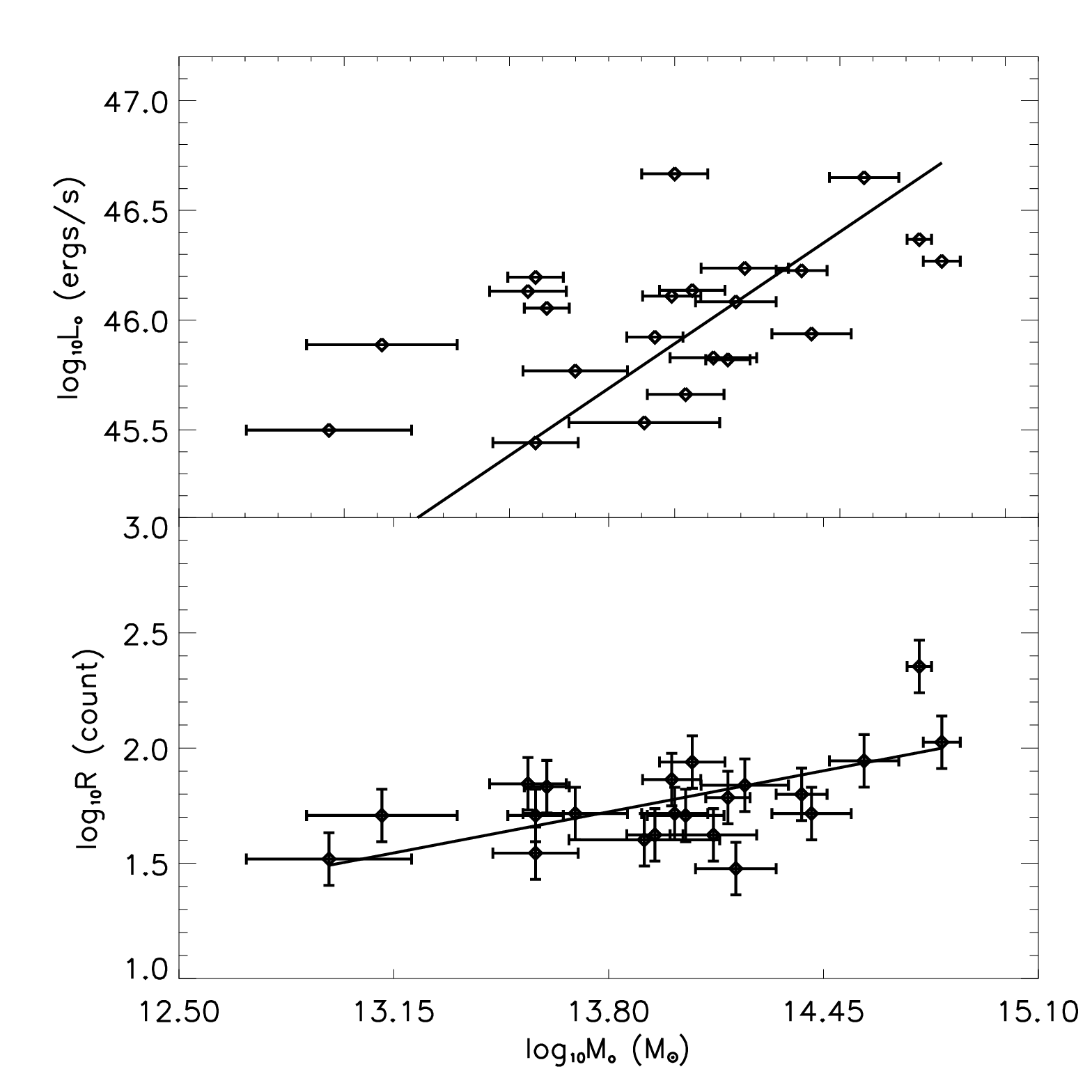}
\caption[]{In the {\bf top} figure, we
plot $L_o$ vs. $M_o$ for the Abell/ACO clusters in the Abell/ACO
sample. We used only those clusters with published virial masses (and
uncertainties) in
Girardi {\it et al.} (1998). The best fit line is $logL_o \propto logM_o^{1.02\pm{0.07}}$.
In the {\bf bottom} figure, we plot Richness
counts (as specified in ACO) vs. $M_o$. The best fit line is 
$logR \propto logM_o^{0.27\pm{0.06}}$. The small slope in $logR$ vs. $logM_o$ makes
correlating mass to richness more difficult than mass to $L_o$.}
\end{figure}
\noindent

{\bf Acknowledgments} ALM thanks Carnegie Mellon University for
support during his sabbatical, as well as the National Center for
Supercomputing Applications for high-speed computing.
We thank Brad Holden for his helpful
discussions. Also, we thank Chris Collins and Stuart Lumsden for
providing unrestricted access to the EDSGC database.  RCN was
partially funded by NASA grant NAG5-3202 during this work.

\begin{deluxetable}{ccc}
\tablenum{1}
\tablewidth{0pt}
\tablecaption{\bf Cluster Luminosities}
\tablehead{
\colhead{Cluster Name} & \colhead{$L_o \times10^{11}$} & \colhead{$L_x \times10^{44}$} \nl
\colhead{} & \colhead{L$\odot$} & \colhead{$h_{50}^{-2}$ergs/s}}
\startdata
EDCC  42 &  $10.04\pm{0.08}$ & $ 5.373\pm{0.69} $ \nl
EDCC 127 &  $ 9.19\pm{0.57}$ & $ 7.753\pm{0.95} $ \nl
EDCC 160 &  $14.85\pm{0.21}$ & $ 9.351\pm{1.05} $ \nl
EDCC 197 &  $ 2.10\pm{0.12}$ & $ 0.270\pm{0.15} $ \nl
EDCC 287 &  $ 5.94\pm{0.68}$ & $ 1.888\pm{1.34} $ \nl
EDCC 320 &  $ 1.73\pm{0.23}$ & $ 0.878\pm{0.41} $ \nl
EDCC 394 &  $ 2.25\pm{0.22}$ & $ 1.157\pm{0.34} $ \nl 
EDCC 400 &  $ 9.88\pm{0.53}$ & $ 2.725\pm{1.11} $ \nl
EDCC 410 &  $ 3.96\pm{0.24}$ & $ 2.881\pm{0.43} $ \nl
EDCC 438 &  $ 3.99\pm{0.07}$ & $ 1.207\pm{0.78} $ \nl
EDCC 447 &  $ 7.12\pm{0.30}$ & $ 2.012\pm{0.75} $ \nl
EDCC 485 &  $ 3.66\pm{0.04}$ & $ 2.373\pm{0.82} $ \nl
EDCC 507 &  $ 9.26\pm{0.62}$ & $ 1.200\pm{0.76} $ \nl
EDCC 520 &  $ 9.68\pm{0.38}$ & $ 3.606\pm{1.63} $ \nl
EDCC 526 &  $ 6.48\pm{0.13}$ & $ 1.338\pm{1.04} $ \nl
EDCC 576 &  $ 2.93\pm{0.15}$ & $ 1.508\pm{1.06} $ \nl
EDCC 632 &  $ 3.05\pm{0.37}$ & $ 1.679\pm{0.52} $ \nl
EDCC 699 &  $ 9.57\pm{0.33}$ & $ 2.556\pm{2.47} $ \nl
EDCC 735 &  $ 5.11\pm{0.13}$ & $ 0.638\pm{0.47} $ \nl
EDCC 758 &  $ 4.77\pm{0.15}$ & $ 1.024\pm{0.32} $ \nl
RASS 001 &  $ 1.84\pm{0.22}$ & - \nl 
RASS 002 &  $10.92\pm{0.05}$ & - \nl 
RASS 003 &  $ 4.22\pm{0.24}$ & - \nl 
RASS 006 &  $ 1.21\pm{0.05}$ & - \nl 
RASS 007 &  $ 3.11\pm{0.24}$ & - \nl 
RASS 008 &  $ 7.18\pm{0.30}$& -  \nl 
RASS 010 &  $ 8.83\pm{0.63}$ & - \nl 
RASS 011 &  $ 6.24\pm{0.44}$ & - \nl 
RASS 015 &  $ 6.74\pm{1.53}$ & - \nl 
RASS 028 &  $ 0.56\pm{0.72}$& -  \nl 
RASS 030 &  $ 3.82\pm{0.84}$ & - \nl 
RASS 106 &  $10.82\pm{0.01}$ & - \nl 
RASS 109 &  $ 8.50\pm{0.07}$ & - \nl 
RASS 110 &  $ 2.64\pm{0.07}$ & - \nl 
RASS 113 &  $ 3.97\pm{0.20}$ & - \nl 
RASS 115 &  $10.70\pm{0.22}$ & - \nl
RASS 127 &  $ 3.76\pm{0.21}$ & - \nl 
RASS 128 &  $ 3.67\pm{0.56}$ & - \nl 
RASS 130 &  $ 4.92\pm{0.74}$& -  \nl 

\enddata
\end{deluxetable}

\end{document}